\documentclass[nopreprintline, final,5p,times,twocolumn]{elsarticle}

\usepackage{amssymb}
\usepackage{graphicx}
\usepackage{physics}
\usepackage{subcaption}
\usepackage{lineno}
\usepackage{wrapfig}
\usepackage{float}
\graphicspath{{./images/}}
\journal{NIMA}

\date{}

\begin{document}
\bibliographystyle{elsarticle-num} 
%\bibliographystyle{abbrvnat} 
%\bibliographystyle{elsarticle-harv} 
%\bibliography{./proceedingrefs.bib}

%\linenumbers

\begin{frontmatter}

\title{Machine Learning based tool for CMS RPC currents quality monitoring}

%\author[1]{Elton Shumka on behalf of the CMS muon group}
%\institute[1]{University of Sofia "St. Kliment Ohridski"}
%\address[1]{University of Sofia "St. Kliment Ohridski"}

%\affil[1]{University of Sofia St.Kliment Ohridski}
%\address[inst1]{organization={University of Sofia "St.Kliment Ohridski"}, addressline={blvd. James Bourchier 8},
%			city={Sofia}, 
%			postcode={1000}
%}
\author[e]{ E. Shumka} 
\ead{elton.shumka@cern.ch}
\author[a]{A. Samalan}
\author[a]{ M. Tytgat}
\author[aa]{ M. El Sawy}
\author[b]{ G.A. Alves}
\author[b]{ F. Marujo}
\author[b]{ E.A. Coelho}
\author[c]{ E.M. Da Costa}
\author[c]{ H. Nogima}
\author[c]{ A. Santoro}
% ...

\address[a]{Ghent University, Dept. of Physics and Astronomy, Proeftuinstraat 86, B-9000 Ghent, Belgium.}
\address[aa]{Université Libre de Bruxelles, Avenue Franklin Roosevelt 50-1050 Bruxelles, Belgium.}
\address[b]{Centro Brasileiro Pesquisas Fisicas, R. Dr. Xavier Sigaud, 150 - Urca, Rio de Janeiro - RJ, 22290-180, Brazil.}
\address[c]{Dep. de Fisica Nuclear e Altas Energias, Instituto de Fisica, Universidade do Estado do Rio de Janeiro, Rua Sao Francisco Xavier, 524, BR - Rio de Janeiro 20559-900, RJ, Brazil.}
% ...
\address[e]{Faculty of Physics, University of Sofia,5 James Bourchier Boulevard, BG-1164 Sofia, Bulgaria.}
% ...
%\pagebreak

%\tableofcontents

\begin{abstract}
%	The muon system of the CMS experiment hosts 1056 Resistive Plate Chambers (RPCs). Detector current monitoring is fundamental for controlling and verifying detector operation. An automated monitoring tool to carry out this task has been developed. It models the behavior of chamber currents by using Machine Learning (ML) methods [1].

The muon system of the CERN Compact Muon Solenoid (CMS) experiment 
includes more than a thousand Resistive Plate Chambers (RPC). They are gaseous detectors operated in the hostile environment of the CMS underground cavern on the Large Hadron Collider where pp luminosities of up to $2\times 10^{34}$ $\text{cm}^{-2}\text{s}^{-1}$ are routinely achieved. The CMS RPC system performance is constantly monitored and the detector is regularly maintained to ensure stable operation. The main monitorable characteristics are dark current, efficiency for muon detection, noise rate etc. Herein we describe an automated tool for CMS RPC current monitoring which uses Machine Learning techniques. We further elaborate on the dedicated generalized linear model proposed already and add autoencoder models for self-consistent predictions as well as hybrid models to allow for RPC current predictions in a distant future.
 
%Two types of ML approaches are used: Generalized Linear Models (GLM) and Autoencoders (AE). 
%In the GLM case, a set of parameters such as environmental conditions, LHC parameters and working point are used to characterize the behavior of the current. \\
%	In the autoencoder case, the set of currents for all of the high-voltage channels of the RPC system are used as input and the autoencoder network is trained to reproduce these inputs on the output neurons. Both approaches show very good predictive capabilities, with accuracy of the order of 1-2 $\mu$A. These predictive capabilities are the basis for the monitoring tool, which is going to be deployed during Run 3. All the developed tools are integrated in a framework that can be easily accessed and controlled by a specially developed Web User Interface that allows the end user to work with the monitoring tool in a simple manner. 
\end{abstract}

\end{frontmatter}

\section{Introduction}
%The muon system of the CMS experiment hosts 1056 Resistive Plate Chambers (RPCs). Detector current monitoring is fundamental for controlling and verifying detector operation. An automated monitoring tool to carry out this task has been developed. It models the behavior of chamber currents by using Machine Learning (ML) methods [1]. 
The muon system of the CMS experiment \cite{CMS} includes 1056 Resistive Plate Chambers (RPC) operated at nominal high voltages (HV) of 9-10 kV.
%The muon system of the CMS experiment hosts 1056 Resistive Plate Chambers  operated at nominal voltages between 9 and 10 kV.  
Monitoring their dark current evolution, spotting 
%deviations from normal performance 
deviations from 
normal performance 
%standard performance 
and anticipating an HV failure that would immediately propagate to higher detector control levels is an unfeasible task for an online operator. Detector parameters are abundant \cite{muonPerformance}, thus HV problems manifest differently, making the human intervention inefficient. 
%Detector parameters are abundant, HV problems manifest differently, human intervention is inefficient.
Therefore, an automated process with built-in notification logic and mechanism is highly sought after for development and further implementation. Being able to spot increasing current tendencies, for example, before they lead to an error is very important for controlling detector operation. An automated tool that performs anomaly detection for the RPC currents by using Machine Learning (ML) methods is presented here. 
\\ \indent Two types of ML approaches are used: Generalized Linear Models (GLM) and Autoencoders.  \\
\indent In the GLM case, a set of parameters such as environmental conditions, LHC parameters and detector working points are used to characterize the behavior of the current. \\
\indent	In the autoencoder case, the full set of the RPC HV system currents is used as an input and the autoencoder network is trained to reproduce these inputs onto the output neurons. \\ \indent Both approaches show very good predictive capabilities that are the basis for the monitoring tool. All the developed tools are integrated in a framework that can be easily accessed and controlled by a specially developed Web User Interface that allows the end-users to work with the monitoring tool in a simple manner. It is being deployed for use during the CERN LHC Run-3 data-taking period.

\section{Generalized Linear Model}
\begin{figure}[h]
	\centering
	\includegraphics[width=1.\linewidth]{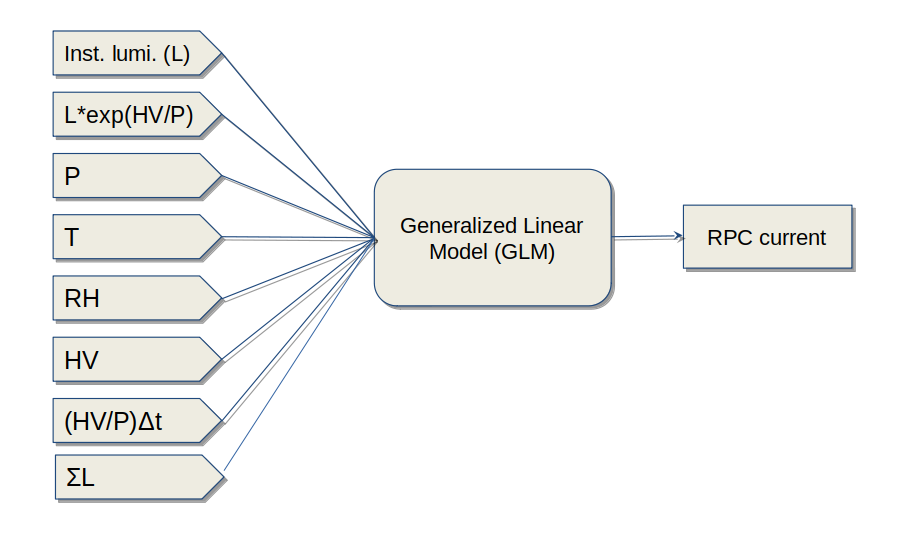}
	\caption{The structure of the GLM}
	\label{fig:glm}
\end{figure}

The GLM depicted in
%The Generalized Linear Model depicted on  
Fig.\,\ref{fig:glm} is a generalization of a simple linear regression
used to model the current as a function of the following sets of parameters: 
\begin{itemize}
	\item Environmental conditions: temperature (T), relative humidity (RH) and pressure (P)
	\item LHC parameters: instantaneous luminosity (L) and integrated luminosity ($\sum$L)
	\item Applied HV
%	\item Applied high-voltage (HV)
	\item Combined terms: $\text{L}\times\text{exp(HV/P)}$ and $(\sum\text{HV}/\text{P})\Delta t$, where $\Delta t$ is the length of the time period with no luminosity
\end{itemize}
The $\sum$L term replaces the $\Delta t$ term used in the initially proposed model \cite{oldpaper}. 
The improvement is inspired by \cite{Gelmi_2021}.
%The improvement was inspired by \cite{Gelmi_2021}.
The first combined term is to account for the exponential increase of gas multiplication with the raising of HV while the second one 
is to account 
%accounts 
for the chamber relaxation and the drop of the current baseline  during cosmic data taking, when there is no beam luminosity and the chambers are at their working point. All the remaining terms and the motivation for including them are discussed in \cite{oldpaper}. 
%\begin{figure}[h]
%	\centering
%	\includegraphics[width=.8\linewidth]{glmv2corr}
%	\caption{The structure of the GLM}
%	\label{fig:glm}
%\end{figure} 

\section{Autoencoder}
\begin{figure}[h]
	\centering
	\includegraphics[width=.9\linewidth]{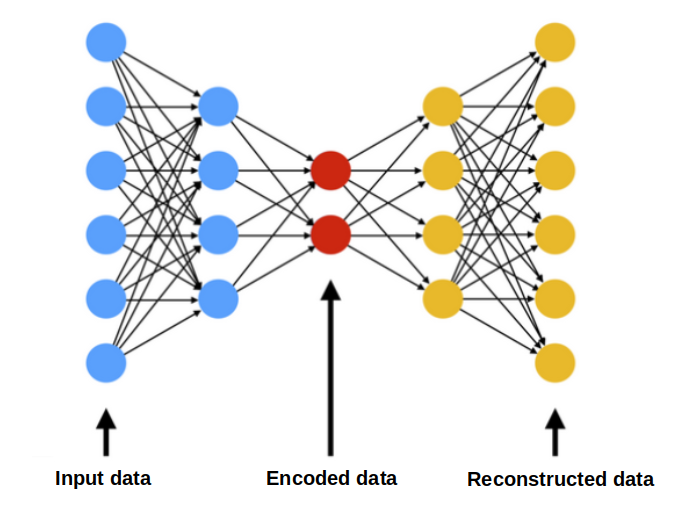}
	\caption{Topology of an autoencoder}
	\label{fig:ae}
\end{figure}

In contrast to the GLM approach, where we use detailed knowledge for the physical processes taking place in a particular type of detector in order to build the ML model, in this section we take a more general approach, namely develop an ML model based on cross-correlation between different detector modules, thus applicable for detector systems consisting of a large number of RPC chambers. We develop an ML algorithm based on an autoencoder model. Autoencoders are neural networks that are trained to encode the input into a number of neurons that is lower than the number of inputs themselves and then decode that same information onto the output layer (Fig.\,\ref{fig:ae}). 
During the learning stage, the autoencoder is supposed to learn the collective behavior of all the RPC chambers. Such an autoencoder could be used later on to spot an anomalous behavior of a single or a small subset of RPC chambers.

%Autoencoders are neural networks that are trained to encode the input in a number of neurons that is fewer than the number of inputs and then decode that same information on the output layer (Fig.  \ref{fig:ae}). \\
%\begin{figure}[h]
%	\centering
%	\includegraphics[width=.7\linewidth]{autoencodercorr}
%	\caption{Topology of an autoencoder}
%	\label{fig:ae}
%\end{figure}

\indent In this work, the set of RPC currents at a given moment in time is given as an input to the autoencoder and the network is trained to reproduce them on its output layer. The number of input and output neurons is 773, which corresponds to the number of HV channels in the RPC system. The hidden layers count respectively: 512, 128, 64, 128 and 512 neurons.

\section{Hybrid network}
\begin{figure}[h]
%	\centering
	\includegraphics[width=1.\linewidth]{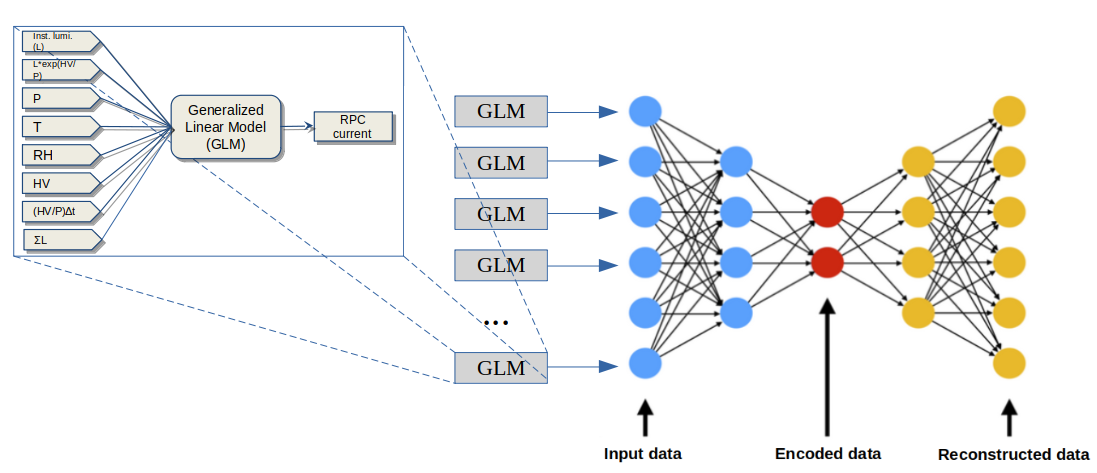}\
	\caption{The hybrid network (the left and right parts of this figure are the same as Figs. \ref{fig:glm} and \ref{fig:ae}, respectively)}
	\label{fig:hybrid}
\end{figure}

As discussed above, GLM describes individual RPC chamber behavior while the autoencoder describes collective correlations of the whole system.
In order to use their best qualities, we combine the two approaches  into a model, referred to as a hybrid network. 
%In the hybrid network, a set of GLM equal in number to the number of HV channels provide as output the currents for a given moment in time.  
In this model, a set of GLM equal in size to the number of HV channels provide as output the currents for a given moment in time.  
%In order to make use of their best qualities, we combine the two approaches  into a model, referred to as a hybrid network. In the hybrid network, a set of GLM equal in number to the number of HV channels provide as output the currents for a given moment in time. 
These currents are then used as inputs for an autoencoder, as shown in Fig.\,\ref{fig:hybrid}.
%\begin{figure}[h]
%%	\centering
%	\includegraphics[width=1.03\linewidth]{hybridnetcorrnew}\
%	\caption{The hybrid network}
%	\label{fig:hybrid}
%\end{figure}
The hybrid network is tested in a distant prediction scenario, where the end of the training period is separated in time (e.g. 1 year) from the beginning of the prediction period. Its performance in such a scenario (Section \ref{sec:results}) shows that it can be used as indication	for current values that we could expect on a system level for some specified conditions (e.g. the luminosity of the High-Luminosity LHC).
%(HL-LHC)).
%(e.g. HL-LHC luminosity).

\section{Monitoring tool}
\begin{figure}[h!]
	\centering
	\includegraphics[width=1.\linewidth]{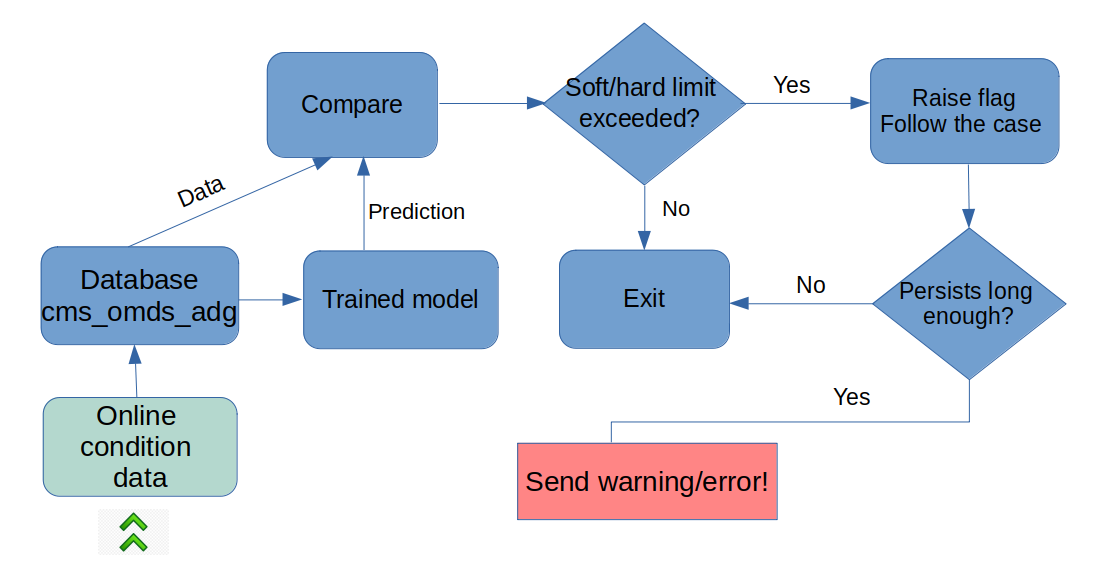}
	\caption{Monitoring tool workflow}
	\label{fig:monit}
\end{figure}

The accurate predictions of the currents performed by both the GLM and autoencoder can be used to detect anomalies in the RPC detector current performance. The implemented tool follows the workflow presented in the flowchart in Fig.\,\ref{fig:monit}.  
%\linebreak
Raw data coming from the CMS non-physics event bus, referred to as online condition data, are written in the $\text{cms}\_\text{omds}\_\text{adg}$ database copy. For each point in time for which data is available, the tool performs comparisons between the measured and predicted RPC currents. 
If differences higher than some predetermined threshold values are detected for a given HV channel, 
%If for a given HV channel differences higher than some predetermined threshold values are detected, 
a flag is raised and the case of that particular channel is followed. There are two thresholds, the lower one inducing a warning and the higher one inducing an error. After a specified number of points in time, the running average of the differences is calculated and if this average exceeds the thresholds, a warning or an error is sent to the end-users. 
This allows for the detection of problematic HV channels before they result in an HV channel trip. 

\section{Software implementation}
\begin{figure}[h]
	\centering
	\includegraphics[width=1.\linewidth]{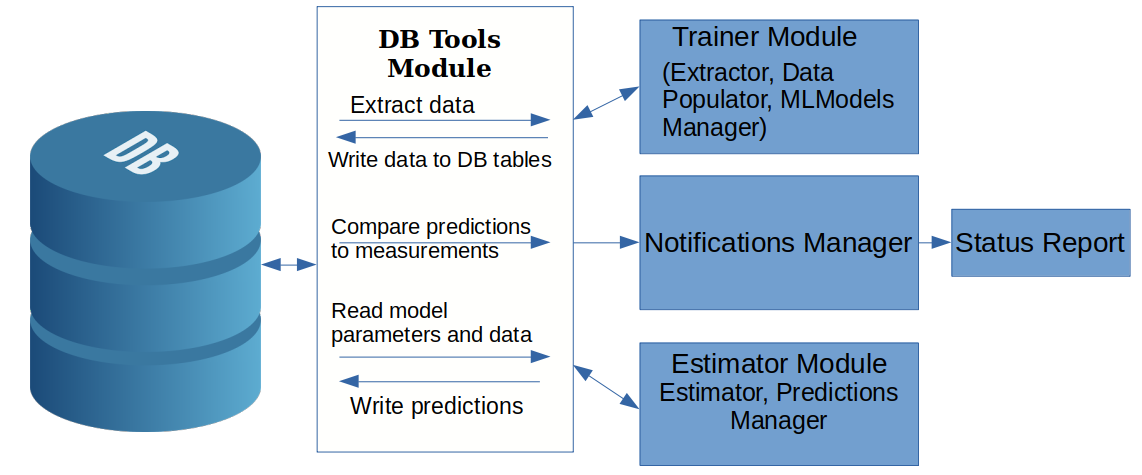}
	\caption{Software structure}
	\label{fig:swstruct}
\end{figure}

The monitoring tool is programmed in Python. Tensorflow \cite{TensorFlow} is used for the implementation of ML. The software is conceptualized and implemented with modularity in mind (Fig.\,\ref{fig:swstruct}).
All modules communicate back-and-forth with a database. The "Trainer Module" reads the training data from a table and after performing the  training, writes back the ML model parameters in another database table. The "Estimator Module" loads the models and performs predictions, which are also stored into the database. Finally, the "Notifications Manager" searches for anomalies in the current values, as described in the previous section and provides notifications.

\begin{figure*}[h!]
\begin{subfigure}{.45\linewidth}
	\centering
	\includegraphics[width=.9\linewidth]{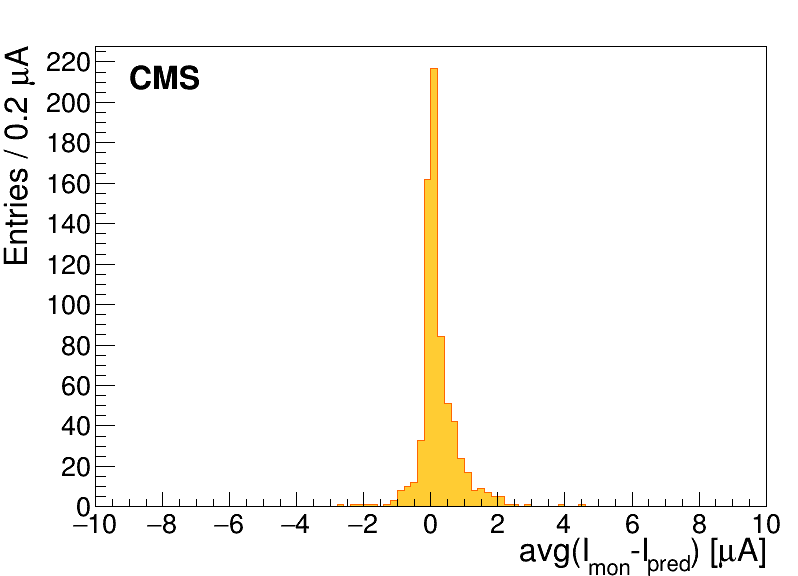}
	\caption{\scriptsize Mean = $0.21$ $\mu$A; $\sigma = 0.59$ $\mu$A}
\end{subfigure}
\begin{subfigure}{.45\linewidth}
	\centering
	\includegraphics[width=.9\linewidth]{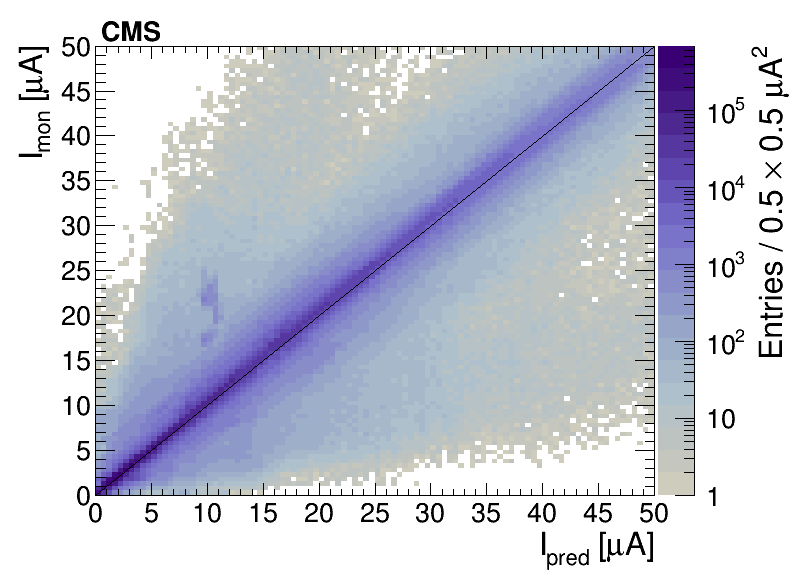}
	\caption{\scriptsize MAE = $0.72$ $\mu$A; MSE = $3.24\mu\text{A}^2$}
\end{subfigure}
\caption{GLM LT performance}
\label{fig:glmv2LT}
\end{figure*}

\begin{figure*}[h!]
\begin{subfigure}{.45\linewidth}
	\centering
	\includegraphics[width=.9\linewidth]{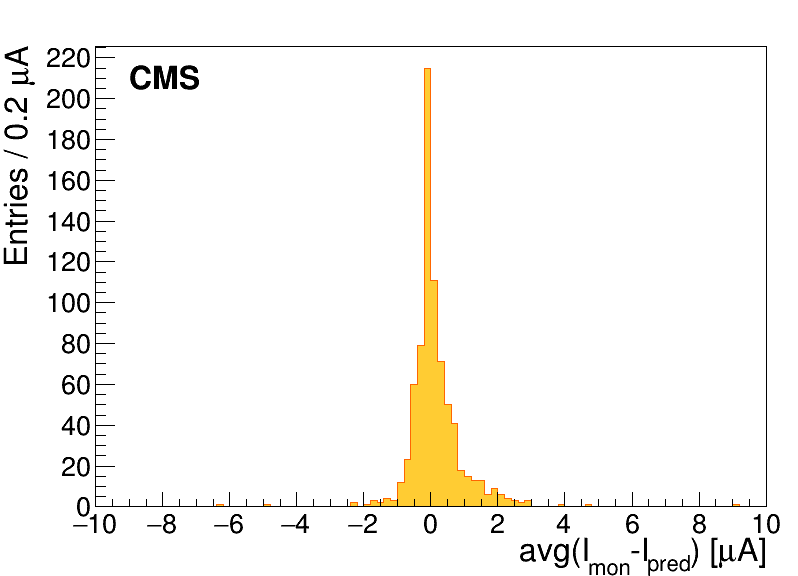}
	\caption{\scriptsize Mean = $0.14$ $\mu$A; $\sigma$ = $0.83$ $\mu$A}
\end{subfigure}
\begin{subfigure}{.45\linewidth}
	\centering
	\includegraphics[width=.9\linewidth]{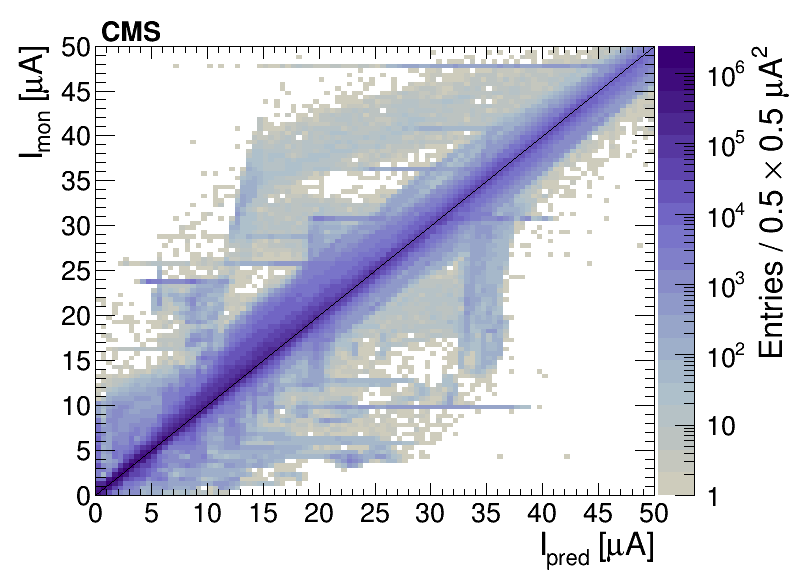}
	\caption{\scriptsize MAE = $0.49$ $\mu$A; MSE = $1.39\mu\text{A}^2$}
\end{subfigure}
\caption{Autoencoder ST performance}
\label{fig:AutoencST}
\end{figure*}

\begin{figure*}[h!]
\begin{subfigure}{.45\linewidth}
	\centering
	\includegraphics[width=.9\linewidth]{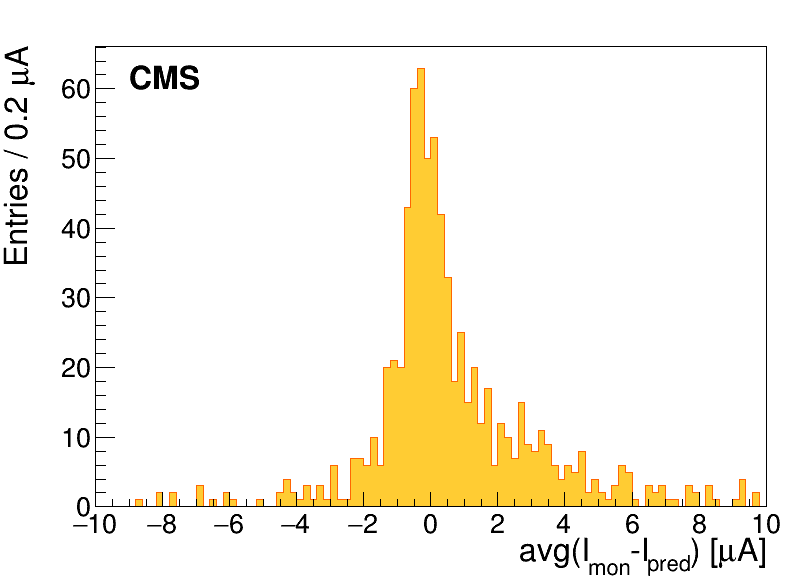}
	\caption{\scriptsize Mean = $0.60$ $\mu$A; $\sigma$ = $2.49$ $\mu$A}
\end{subfigure}
\begin{subfigure}{.45\linewidth}
	\centering
	\includegraphics[width=.9\linewidth]{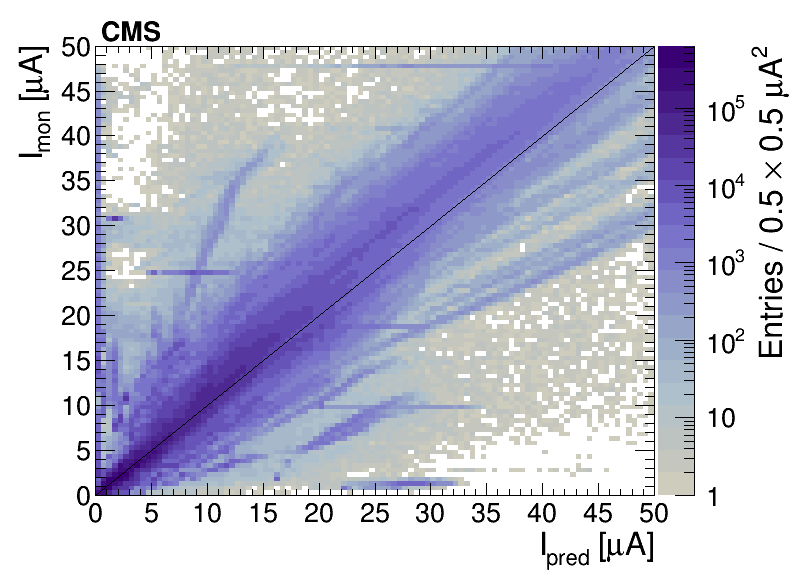}
	\caption{\scriptsize MAE = $2.09$ $\mu$A; MSE = $23.19\mu\text{A}^2$}\end{subfigure}
\caption{Hybrid network performance}
\label{fig:Hybrid}
\end{figure*}

\section{Performance results}
\label{sec:results}
ML model performance validation is done for three different training scenarios:
\begin{itemize}
	\item Short-term training (ST), with data from May to September 2018. Such models are able to spot a rapid increase in the RPC currents.
	\item Mid-term training (MT), with data from July 2017 to July 2018, appropriate for describing the seasonal behavior of the currents.
	\item Long-term training (LT), with data from May 2016 to July 2018, appropriate for modelling the overall RPC currents evolution. 
\end{itemize}

All models are tested against the RPC currents measured in the two-month period between September and October of 2018. 
%All models are tested against the measured RPC currents from September to the end of October 2018. 
These tests show that GLM performs best in LT scenario (Fig.\,\ref{fig:glmv2LT}), while the autoencoder performs best in ST (Fig.\,\ref{fig:AutoencST}). The Mean Absolute Error (MAE) and Mean Squared Error (MSE), which are used as performance metrics,  are defined as:
\begin{equation}
	\text{MAE} = \displaystyle \sum_{i=1}^N \frac{|\text{I}_{mon}^i - 		\text{I}_{pred}^i|}{\text{N}}
	\label{eq:mae}
\end{equation}

\begin{table*}[h]
\centering
\caption{Performance results}
\scriptsize
\begin{tabular}{|p{0.1\linewidth}|p{0.12\linewidth}|p{0.12\linewidth}|p{0.12\linewidth}|p{0.12\linewidth}|p{0.12\linewidth}|p{0.12\linewidth}|}
%\begin{tabular}{|c|c|c|c|c|c|c|}
\hline
	Model class & Training period & Prediction period & 1D histo mean [$\mu\text{A}$] & 1D histo sigma [$\mu\text{A}$] & 2D histo MAE [$\mu\text{A}$] & 2D histo MSE [$\mu\text{A}^2$]\\
\hline 
	GLMv2 & 18-05-01 to 18-09-01 & 18-09-01 to 18-10-30 & -0.02 & 1.65 & 1.23 & 7.62 \\
\hline
	GLMv2 & 17-07-01 to 18-07-01 & 18-09-01 to 18-10-30 & 0.33 & 1.66 & 1.23 & 7.42 \\
\hline 
	GLMv2 & 16-05-01 to 18-07-01 & 18-09-01 to 18-10-30 & 0.21 & 0.59 & 0.72 & 3.24 \\
\hline
	Autoencoder & 18-05-01 to 18-09-01 & 18-09-01 to 18-10-30 & 0.14 & 0.83 & 0.49 & 1.39 \\
\hline
	Autoencoder & 17-07-01 to 18-07-01 & 18-09-01 to 18-10-30 & 0.69 & 1.44 & 0.96 & 4.18 \\
\hline
    Autoencoder & 16-05-01 to 18-07-01 & 18-09-01 to 18-10-30 & 0.42 & 1.40 & 0.85 & 3.16 \\ 
\hline 
	GLMv2 & 16-05-01 to 17-07-01 & 18-09-01 to 18-09-30 & -0.24 & 2.59 & 1.92 & 18.69 \\ 
\hline 
	Autoencoder & 16-05-01 to 17-07-01 & 18-09-01 to 18-09-30 &  0.06 & 2.51 & 2.14 & 22.57 \\ 
\hline 
	Hybrid & 16-05-01 to 17-07-01 & 18-09-01 to 18-09-30 & 0.60 & 2.49 & 2.09 & 23.19 \\
\hline
\end{tabular}
\label{tab:results}
\end{table*}

%\begin{figure*}[h]
%	\centering
%	\includegraphics[width=.75\linewidth]{example1}
%	\caption{Monitored and predicted currents for an RPC chamber in W-1 of the CMS barrel}
%	\label{fig:example}
%\end{figure*}

\begin{figure*}[h!]
	\centering
	\includegraphics[width=.95\linewidth]{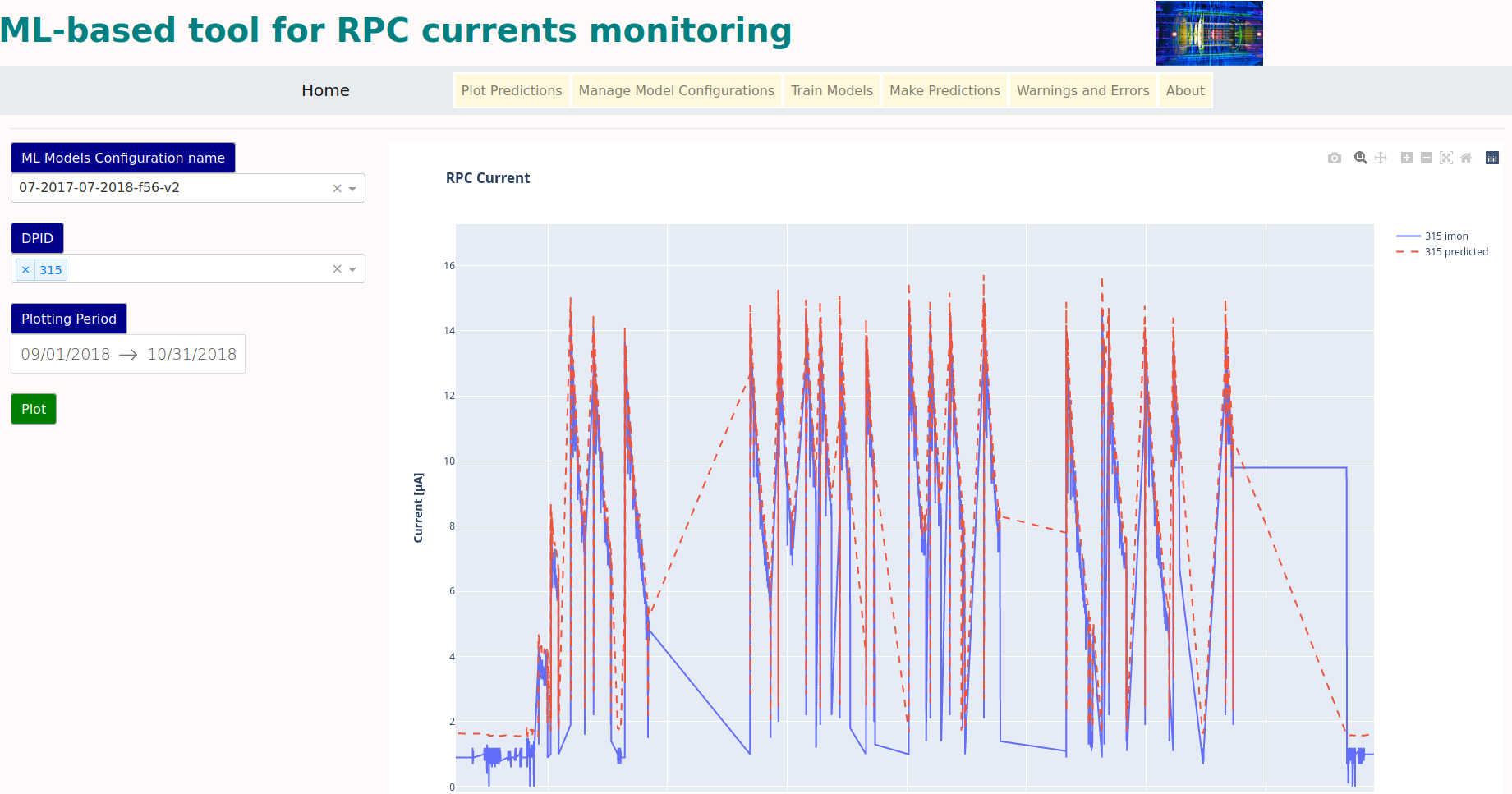}
	\caption{Screenshot of the Web User Interface}
	\label{fig:ui}
\end{figure*}
\begin{figure}[h]
	\centering
	\includegraphics[width=.95\linewidth]{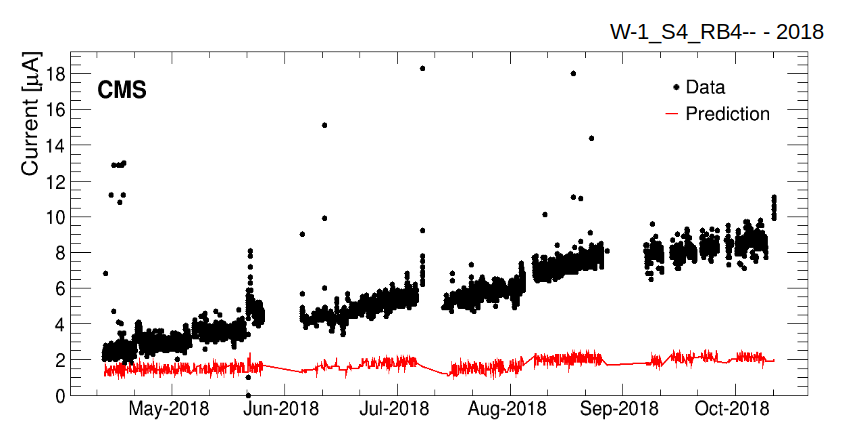}
	\caption{Monitored and predicted currents for an RPC chamber in W-1 of the CMS barrel}
	\label{fig:example}
\end{figure}


\begin{equation}
		\text{MSE} = \displaystyle \sum_{i=1}^N \frac{(\text{I}_{mon}^i - 		\text{I}_{pred}^i)^2}{\text{N}}
		\label{eq:mse}
\end{equation}
\\
The sigma of the histogram in both cases is $< 1 \mu A$, which shows that both models have excellent predictive capabilities consistent with the uncertainty in the current measurement which is also of that same order. 
All performance results are shown in Table \ref{tab:results}. \\
\indent Fig. \ref{fig:example} shows the case of an RPC chamber where the predicted current increasingly diverges from the measured one. It was found that this discrepancy could be explained with the appearance of a gas leak in this chamber around the same time.

\section{Deployment on the CERN PaaS platform}
The monitoring tool is accessible through a Web User Interface that is being deployed on the CERN Platform-as-a-Service (PaaS) virtual environment (Fig. \ref{fig:ui}). It is based on OpenShift \cite{openshift}, a platform which allows for containerized application deployment.

%\begin{figure*}[h]
%	\centering
%	\includegraphics[width=.9\linewidth]{uipred}
%	\caption{Screenshot of the Web User Interface}
%	\label{fig:ui}
%\end{figure*}

%\begin{figure}[h]
%	\centering
%	\includegraphics[width=1.4\linewidth, angle=270]{uipred}
%	\caption{Screenshot of the Web User Interface}
%	\label{fig:ui}
%\end{figure}

\section{Conclusions}
We use Machine Learning (ML) methods for anomaly detection in the current behavior of CMS Resistive Plate Chambers. 
%CMS Resistive Plate Chambers current behavior. 
The excellent accuracy of the ML model predictions allow us to implement a powerful monitoring tool which notifies the end-users about potential high-voltage channel deviations from 
normal behavior 
%standard behavior 
and increased risk of operational failures.
The monitoring tool has been developed and will be fully deployed for use during the Year-End Technical Stop (YETS22/23). 

\section*{Acknowledgments}
We congratulate our colleagues in the CERN accelerator departments for the excellent performance of the LHC and thank the technical and administrative staffs at CERN and at other CMS institutes for their contributions to the success of the CMS effort. In addition, we gratefully acknowledge the computing centres and personnel of the Worldwide LHC Computing Grid and other centres for delivering so effectively the computing infrastructure essential to our analyses. Finally, we acknowledge the enduring support for the construction and operation of the LHC, the CMS detector, and the supporting computing infrastructure provided by the following funding agencies: FWO (Belgium); CNPq, CAPES and FAPERJ (Brazil); MES and BNSF (Bulgaria); CERN; CAS, MoST, and NSFC (China); MINCIENCIAS (Colombia); CEA and CNRS/IN2P3 (France); SRNSFG (Georgia); DAE and DST (India); IPM (Iran); INFN (Italy); MSIP and NRF (Republic of Korea); BUAP, CINVESTAV, CONACYT, LNS, SEP, and UASLP-FAI (Mexico); PAEC (Pakistan); DOE and NSF (USA). 
%\printbibliography
%\bibliographystyle{elsarticle-harv} 
\bibliography{proceedingrefs}

%\begin{thebibliography}{00}
%
%\bibitem{CMS} CMS Collaboration, "The CMS Experiment at the CERN LHC", JINST 3 (2008) S08004, \href{https://doi.org/10.1088/1748-0221/3/08/
%
%\bibitem{oldpaper} A. Salaman et al., A new approach for CMS RPC current monitoring using Machine Learning techniques, JINST 15 (2020) 10, C10009, DOI:10.1088/1748-0221/15/10/C10009
%S08004}{DOI:10.1088/1748-0221/3/08/S08004}
%
%\bibitem{CMSmuonperformance} CMS Collaboration, Performance of the CMS muon detector and muon reconstruction with proton-proton collisions at s√= 13 TeV, JINST 13 (2018) P06015, https://doi.org/10.48550/arXiv.1804.04528
%
%\bibitem{TensorFlow} \href{https://www.tensorflow.org/}{TensorFlow}
%
%\end{thebibliography}

\end{document}